\documentclass[11pt,a4paper]{article}
\pdfoutput=1
\usepackage{jheppub}
\usepackage{amsmath,amssymb}
\usepackage{graphicx}
\usepackage{bm}
\usepackage{xcolor}
\usepackage{booktabs}
\usepackage{xcolor}

\title{Generalized Virial Identities:\\ Radial Constraints for Solitons, Instantons, and Bounces}

\author[a]{Jonathan Lozano-Mayo}
\affiliation[a]{Weinberg Institute for Theoretical Physics, Department of Physics,\\
The University of Texas at Austin, Austin, TX 78712, USA}
\emailAdd{jonathanloz@utexas.edu}

\abstract{ We derive a continuous family of virial identities for O($n$) symmetric configurations, parameterized by an exponent $\alpha$ that controls the radial weighting. The family provides a systematic decomposition of the global constraint into radially-resolved components, with special $\alpha$ values isolating specific mechanisms. For BPS configurations, where the Bogomolny equations imply pointwise equality between kinetic and potential densities, the virial identity is satisfied for all valid $\alpha$. We verify the formalism analytically for the Fubini-Lipatov instanton, BPS monopole, and BPST instanton. Numerical tests on the Coleman bounce and Nielsen-Olesen vortex illustrate how the $\alpha$-dependence of errors distinguishes core from tail inaccuracies: the vortex shows errors growing at negative $\alpha$ (core), while the bounce shows errors growing at positive $\alpha$ (tail). Applications to the electroweak sphaleron, where the Higgs mass explicitly breaks scale invariance, and the hedgehog Skyrmion illustrate the formalism in systems with multiple competing length scales.}
\keywords{Solitons Monopoles and Instantons, Nonperturbative Effects, Effective Field Theories}
\begin{document}
\maketitle

\section{Introduction}
\label{sec:intro}

Topological solitons and tunneling configurations occupy a central place in modern field theory~\cite{MantonSutcliffe2004,Shnir2005}. Magnetic monopoles~\cite{tHooft1974,Polyakov1974} arise as inevitable consequences of grand unified symmetry breaking and carry implications for early-universe cosmology. Cosmic strings and vortices~\cite{NielsenOlesen1973,Vilenkin1985} seed density perturbations and produce distinctive gravitational wave signatures. The electroweak sphaleron~\cite{KlinkhamerManton1984,MatchevVerner2025} mediates baryon number violation at high temperatures, while Skyrmions~\cite{Skyrme1962,AdkinsNappiWitten1983,MantonSkyrmions2022} model baryons in the large-$N_c$ limit of QCD. Bounce solutions~\cite{Coleman1977,ColemanCallan1977,ColemanDeLuccia1980} govern vacuum decay rates in scalar field theories and play an essential role in discussions of Higgs metastability. These configurations share a common mathematical structure: they extremize energy or action functionals subject to topological or boundary constraints, and their properties (mass, size, interaction potential) are encoded in the profile functions that describe them.

While a handful of these solutions admit closed-form expressions (the BPS monopole, the BPST instanton, the Fubini-Lipatov instanton), most must be obtained numerically. The question we address in this paper is not how to obtain such solutions, but rather how to characterize their radial structure through integral identities that encode the underlying physics. The most familiar such identity is Derrick's theorem~\cite{Derrick1964}, which provides a single integral constraint relating total kinetic and potential contributions. Under a uniform dilation $x \to \lambda x$, these contributions scale differently, and stationarity at $\lambda = 1$ yields a relation that any equilibrium configuration must satisfy. This relation has profound consequences: it forbids stable scalar solitons in three or more spatial dimensions when the potential is positive-definite, and it explains why the Skyrme term (quartic in derivatives) is necessary for Skyrmion stability. However, the Derrick relation is a global constraint that integrates over all radii and reveals nothing about the local balance between kinetic and potential contributions in different regions of the soliton.

Our central observation is that the geometric factor $\rho^{n-1}$ appearing in the O($n$) symmetric measure introduces explicit $\rho$-dependence into the reduced one-dimensional problem, and this dependence can be exploited systematically. By weighting the fundamental virial relation with $\rho^\alpha$ before integration, we obtain a continuous family of identities indexed by $\alpha$. Negative $\alpha$ emphasizes the core region, where topological boundary conditions are imposed and field gradients are steepest; large positive $\alpha$ emphasizes the asymptotic tail, where the fields approach their vacuum values; and $\alpha = 1$ recovers the classical Derrick relation. The $\alpha$-family thus gives a decomposition of the global Derrick constraint into radially-resolved components, each able to probe different regions of the soliton.

This decomposition has both theoretical and practical value. On the theoretical side, special $\alpha$ values isolate specific stabilization mechanisms. For the Skyrmion, $\alpha = 0$ probes the balance between Dirichlet gradient energy and the centrifugal barrier in the core, while $\alpha = 2$ probes the intermediate region where the Skyrme and sigma-model terms compete. For the electroweak sphaleron, different $\alpha$ values separate the contributions from gauge kinetic energy, Higgs gradients, and vacuum compression. On the practical side, the $\alpha$-family provides independent checks on numerical solutions: a solution that satisfies the $\alpha = 1$ relation but violates $\alpha = -0.5$ has errors concentrated in the core, while one failing at large $\alpha$ has errors in the tail. We illustrate this explicitly for Nielsen-Olesen vortices, where the $\alpha = 1$ identity holds to $0.0005\%$ while $\alpha = -0.5$ shows a $5.7\%$ discrepancy, exposing core-region inaccuracies that the global Derrick test averages away. For the Coleman bounce, the opposite pattern emerges: errors grow with $\alpha$ because the field approaches the false vacuum asymptotically and numerical truncation of the domain affects the tail most.

For BPS configurations, the situation simplifies considerably. The first-order Bogomolny equations imply pointwise equality between kinetic and potential densities, not merely integral equality. When integrated with any weight $\rho^\alpha$, these local identities guarantee automatic satisfaction of the virial identity for all valid $\alpha$. BPS configurations saturate a topological energy bound, and the first-order equations express the condition for saturation; this condition is local and therefore survives integration with any weight. By contrast, non-BPS configurations satisfy only second-order Euler-Lagrange equations, which constrain the total variation of the action but do not imply pointwise balances. The $\alpha$-family for non-BPS solutions thus yields genuinely independent constraints at each $\alpha$.

Our work connects to several recent developments in integral identities for solitons. Manton~\cite{Manton2009} derived scaling identities beyond Derrick's theorem by considering more general coordinate transformations, applying infinitesimal diffeomorphisms to the action functional. Our $\alpha$-family overlaps with Manton's identities at specific parameter values but differs in emphasis: we parameterize the identities continuously and focus on their utility as radial probes. Adam \emph{et al.}~\cite{Adam2024} demonstrated that for any nonlinear field theory supporting static solutions with finite energy, an infinite number of integral identities can be derived from field transformations and their Noether currents. They showed that identities generated by coordinate transformations become trivial for BPS solitons, while field-space transformations yield nontrivial relations connected to observables such as the $D$-term and mechanical radius~\cite{PolyakovSchweitzer2018,Ji2021}. Our formalism provides explicit realizations of coordinate-based identities for O($n$) symmetric configurations and confirms the BPS triviality. Gudnason, Gao, and Yang~\cite{GudnasonGaoYang2017} established a related 3-parameter family of integral identities for spherically symmetric solitons using boundary charge methods; our approach differs in deriving the identities from the radial functional structure. Herdeiro \emph{et al.}~\cite{HerdeiroOliveiraPomboRadu2021,HerdeiroOliveiraPomboRadu2022} extended virial identities to relativistic gravity, showing how they can be recast as energy-momentum balance conditions for self-gravitating solitons and black holes. The $\alpha$-family also generalizes the classical Pohozaev identities~\cite{Pohozaev1965} from elliptic PDE theory: Pohozaev derived integral constraints by multiplying the equation $-\Delta u = f(u)$ by $x \cdot \nabla u$ and integrating by parts, which for radial solutions corresponds to $\alpha = 1$. Our generalization to arbitrary $\alpha$ yields a continuous family of constraints.

The paper is organized as follows. Section~\ref{sec:formalism} develops the general formalism, deriving the virial identities for both scalar and gauge theories. Section~\ref{sec:scalar} applies the formalism to scalar field configurations: the Fubini-Lipatov instanton and the Coleman bounce. Section~\ref{sec:gauge} treats gauge theories: the BPS monopole, the BPST instanton, Nielsen-Olesen vortices, and the electroweak sphaleron. Section~\ref{sec:chiral} applies the formalism to the hedgehog Skyrmion, where the interplay between radial and angular strain determines the profile. We conclude in Section~\ref{sec:conclusion}.

\section{General Formalism}
\label{sec:formalism}

\subsection{Radial  Functional Density}

We consider functionals of the form $F[\phi] = \int d^n x \, \mathcal{G}(\phi, \partial\phi)$ that admit O($n$) symmetric solutions depending only on $\rho = |x|$. This framework encompasses Euclidean instantons (where $F = S_E$ and $n$ is the spacetime dimension) and static solitons (where $F = E$ and $n$ is the number of spatial dimensions).

For O($n$) symmetric configurations, angular integration yields the reduced functional
\begin{equation}
F = \Omega_{n-1} \int_0^\infty \mathcal{G}_\rho \, d\rho,
\label{eq:radial_reduction}
\end{equation}
where $\Omega_{n-1} = 2\pi^{n/2}/\Gamma(n/2)$ is the area of the unit $(n-1)$-sphere and $\mathcal{G}_\rho$ is the reduced integrand. The geometric factor $\rho^{n-1}$ from the volume element $d^n x = \rho^{n-1} d\rho\, d\Omega_{n-1}$ introduces explicit $\rho$-dependence into $\mathcal{G}_\rho$, even when the original functional density $\mathcal{G}$ has no such dependence. This explicit $\rho$-dependence is the origin of all virial constraints.

To see this, consider what would happen if $\mathcal{G}_\rho$ had no explicit dependence on $\rho$. The reduced problem would then possess a continuous symmetry under ``radial translations'' $\rho \to \rho + \epsilon$, and Noether's theorem would yield a conserved quantity. The geometric weights from the integration measure break this would-be symmetry, and the virial identities quantify precisely how it is broken.

The virial identities follow from the Euler-Lagrange equations of the reduced problem. For a reduced integrand $\mathcal{G}_\rho(\rho, \{\phi_a\}, \{\dot{\phi}_a\})$ depending on multiple fields $\phi_a$ with radial derivatives $\dot{\phi}_a = d\phi_a/d\rho$, the Euler-Lagrange equations are
\begin{equation}
\frac{d}{d\rho}\left(\frac{\partial \mathcal{G}_\rho}{\partial \dot{\phi}_a}\right) = \frac{\partial \mathcal{G}_\rho}{\partial \phi_a}, \qquad a = 1, \ldots, N.
\label{eq:EL}
\end{equation}
We define the auxiliary quantity
\begin{equation}
\mathcal{C}_\rho \equiv \sum_a \dot{\phi}_a\frac{\partial \mathcal{G}_\rho}{\partial \dot{\phi}_a} - \mathcal{G}_\rho,
\label{eq:C_general}
\end{equation}
which is the Legendre transform of $\mathcal{G}_\rho$ with respect to the velocities. Computing its derivative along a solution:
\begin{align}
\frac{d\mathcal{C}_\rho}{d\rho} &= \sum_a \left[\ddot{\phi}_a \frac{\partial \mathcal{G}_\rho}{\partial \dot{\phi}_a} + \dot{\phi}_a \frac{d}{d\rho}\left(\frac{\partial \mathcal{G}_\rho}{\partial \dot{\phi}_a}\right)\right] - \frac{d\mathcal{G}_\rho}{d\rho} \nonumber \\
&= \sum_a \left[\ddot{\phi}_a \frac{\partial \mathcal{G}_\rho}{\partial \dot{\phi}_a} + \dot{\phi}_a \frac{\partial \mathcal{G}_\rho}{\partial \phi_a}\right] - \frac{\partial \mathcal{G}_\rho}{\partial \rho} - \sum_a \left[\frac{\partial \mathcal{G}_\rho}{\partial \phi_a}\dot{\phi}_a + \frac{\partial \mathcal{G}_\rho}{\partial \dot{\phi}_a}\ddot{\phi}_a\right],
\label{eq:C_deriv}
\end{align}
where in the second line we used the Euler-Lagrange equations~\eqref{eq:EL} and expanded the total derivative of $\mathcal{G}_\rho$. The terms involving $\ddot{\phi}_a$ and $\dot{\phi}_a$ cancel exactly, leaving the fundamental identity
\begin{equation}
\frac{d\mathcal{C}_\rho}{d\rho} = -\frac{\partial \mathcal{G}_\rho}{\partial \rho}.
\label{eq:C_evolution_general}
\end{equation}
The rate of change of the Legendre transform of the functional $\mathcal{C_\rho}$ equals (minus) the explicit $\rho$-derivative of the functional. If $\mathcal{G}_\rho$ had no explicit $\rho$-dependence, $\mathcal{C}_\rho$ would be conserved, the analogue of energy conservation in time-independent mechanics. The geometric weights break this conservation, and Eq.~\eqref{eq:C_evolution_general} quantifies the breaking.

\subsection{$\alpha$-Family of Virial identities}

We now exploit Eq.~\eqref{eq:C_evolution_general} to generate a continuous family of integral constraints. Multiplying both sides by $\rho^\alpha$ and integrating from $0$ to $\infty$:
\begin{equation}
\int_0^\infty \rho^\alpha \frac{d\mathcal{C}_\rho}{d\rho}\, d\rho = -\int_0^\infty \rho^\alpha \frac{\partial \mathcal{G}_\rho}{\partial \rho}\, d\rho.
\label{eq:weighted_int}
\end{equation}
Integrating the left-hand side by parts:
\begin{equation}
\int_0^\infty \rho^\alpha \frac{d\mathcal{C}_\rho}{d\rho}\, d\rho = \left[\rho^\alpha \mathcal{C}_\rho\right]_0^\infty - \alpha \int_0^\infty \rho^{\alpha-1} \mathcal{C}_\rho\, d\rho.
\label{eq:parts}
\end{equation}
Provided the boundary term $[\rho^\alpha \mathcal{C}_\rho]_0^\infty$ vanishes and the integrals converge (conditions analyzed in Section~\ref{sec:boundary}), we obtain the $\alpha$-family of virial identities:
\begin{equation}
\alpha \int_0^\infty \rho^{\alpha-1} \mathcal{C}_\rho\, d\rho = \int_0^\infty \rho^\alpha \frac{\partial \mathcal{G}_\rho}{\partial \rho}\, d\rho.
\label{eq:alpha_virial}
\end{equation}
This identity holds for each value of $\alpha$ for which these conditions are satisfied. Each identity differs because the weighting function $\rho^\alpha$ emphasizes different radial regions: small (or negative) $\alpha$ weights the core, large $\alpha$ weights the tail, and $\alpha = 1$ provides uniform weighting. By varying $\alpha$, we scan through the radial profile, obtaining independent constraints that collectively characterize the solution's structure. Nothing restricts us to the weight $\rho^\alpha$. Any well-behaved function $f(\rho)$ multiplying Eq.~\eqref{eq:C_evolution_general} yields a valid identity. We focus on $\rho^\alpha$ because the single parameter $\alpha$ provides a clean interpolation between core and tail.

\subsection{Kinetic-potential decomposition}

The identity~\eqref{eq:alpha_virial} takes a more transparent form when $\mathcal{G}_\rho$ separates into kinetic and potential contributions:
\begin{equation}
\mathcal{G}_\rho = \sum_i A_i(\rho) T_i(\{\dot{\phi}_a\}) + \sum_j B_j(\rho) U_j(\{\phi_a\}),
\label{eq:general_structure}
\end{equation}
where $T_i$ are kinetic terms (homogeneous of degree 2 in the velocities $\dot{\phi}_a$) and $U_j$ constitute the effective potential of the reduced system. The geometric weights $A_i(\rho)$, $B_j(\rho)$ encode dimensional factors from the integration measure and may differ between terms, a feature that becomes important in gauge theories where different fields carry different angular momentum.

The kinetic terms $T_i$ represent the ``gradient pressure'' of the field, i.e., the energetic cost of spatial variation. The potential terms $U_j$ represent ``confinement'' or ``binding,'' the energetic cost of the field deviating from its vacuum value. A localized solution exists when these competing effects balance: the gradient pressure resists collapse while the potential prevents dispersal. The virial identities constrain this competition at each radial scale.

By Euler's theorem for homogeneous functions, $\sum_a \dot{\phi}_a \partial T_i / \partial \dot{\phi}_a = 2T_i$ for terms quadratic in velocities. The auxiliary quantity then becomes
\begin{equation}
\mathcal{C}_\rho = \sum_i A_i(\rho) T_i - \sum_j B_j(\rho) U_j.
\label{eq:C_KV}
\end{equation}
Thus $\mathcal{C}_\rho$ is the difference between (weighted) kinetic and potential contributions, precisely the quantity that measures the local kinetic-potential competition. In regions where kinetic energy dominates, $\mathcal{C}_\rho > 0$; in regions where potential energy dominates, $\mathcal{C}_\rho < 0$. The virial identities constrain the $\alpha$-weighted integral of this local imbalance.

The explicit $\rho$-derivative of $\mathcal{G}_\rho$ involves only the weight functions:
\begin{equation}
\frac{\partial \mathcal{G}_\rho}{\partial \rho} = \sum_i \dot{A}_i(\rho) T_i + \sum_j \dot{B}_j(\rho) U_j,
\end{equation}
where $\dot{A}_i = dA_i/d\rho$. Substituting into Eq.~\eqref{eq:alpha_virial} and rearranging yields the explicit form
\begin{equation}
\sum_i \int_0^\infty T_i \left(\alpha \rho^{\alpha-1} A_i - \rho^\alpha \dot{A}_i\right) d\rho = \sum_j \int_0^\infty U_j \left(\alpha \rho^{\alpha-1} B_j + \rho^\alpha \dot{B}_j\right) d\rho.
\label{eq:general_virial}
\end{equation}
The coefficients multiplying $T_i$ and $U_j$ are determined entirely by the geometric structure of the problem.

The procedure for applying this formalism to a specific system is: (i) write the full Lagrangian or energy functional of the field theory; (ii) specify the O($n$) symmetric ansatz that reduces the system to a radial problem; (iii) perform the angular integration to obtain the reduced functional $\mathcal{G}_\rho$; (iv) compute the auxiliary quantity $\mathcal{C}_\rho$ from~\eqref{eq:C_general}; (v) apply the fundamental identity~\eqref{eq:C_evolution_general} and integrate with weight $\rho^\alpha$ to obtain the virial constraints. In subsequent sections, we carry out this procedure explicitly for scalar instantons, gauge field configurations, and chiral solitons.

\subsection{Scalar fields}

For a single scalar field in $n$ dimensions with standard kinetic term $\frac{1}{2}(\nabla\phi)^2$ and potential $V(\phi)$, both weights equal the geometric measure: $A = B = \rho^{n-1}$. The gradient pressure and potential confinement thus carry identical geometric weights, and the general identity~\eqref{eq:general_virial} reduces to
\begin{equation}
(\alpha - n + 1) \int_0^\infty \frac{1}{2}\dot{\phi}^2 \rho^{\alpha+n-2}\, d\rho = (\alpha + n - 1) \int_0^\infty V(\phi) \rho^{\alpha+n-2}\, d\rho,
\label{eq:scalar_virial}
\end{equation}
or equivalently
\begin{equation}
\frac{I_T^\alpha}{I_V^\alpha} = \frac{\alpha + n - 1}{\alpha - n + 1},
\label{eq:ratio}
\end{equation}
where $I_T^\alpha$ and $I_V^\alpha$ are the $\alpha$-weighted kinetic and potential moments.

Equation~\eqref{eq:ratio} simplifies at specific $\alpha$ values. At $\alpha = n - 1$, the kinetic coefficient vanishes: which requires $I_V^\alpha = 0$. This can only hold if the potential is indefinite (taking both positive and negative values along the profile) or if the solution is trivial. At $\alpha = -(n-1)$, the potential coefficient vanishes, requiring $I_T^\alpha = 0$, which is impossible for a nontrivial profile. These singular values mark the boundaries of the allowed $\alpha$ range.

\subsection{Boundary conditions}
\label{sec:boundary}

The virial identity~\eqref{eq:general_virial} holds if and only if two conditions are satisfied: the boundary term $[\rho^\alpha \mathcal{C}_\rho]_0^\infty$ vanishes, and the integrals on both sides converge. Together, these conditions determine the range of $\alpha$ for which the identity is valid, and this range encodes information about the solution's behavior at the core and tail.

At large $\rho$, localized solutions approach their vacuum configuration. For massive theories with exponential decay $\phi - \phi_v \sim e^{-m\rho}$, the boundary term vanishes and integrals converge for all finite $\alpha$, since exponential suppression defeats any polynomial growth. For power-law decay $\phi - \phi_v \sim \rho^{-p}$ (characteristic of massless or conformally coupled theories), convergence requires $\alpha < 2p + 3 - n$; beyond this threshold, the $\rho^\alpha$ weight amplifies the slowly decaying tail faster than the kinetic terms can suppress it.

At the origin, regularity of the solution typically requires $\dot{\phi}(0) = 0$ (the field approaches the core as a local extremum). The kinetic integral then converges near $\rho = 0$ when $\alpha > 1 - n$; below this threshold, the $\rho^{\alpha-1}$ factor in the kinetic integrand diverges faster than the $\dot{\phi}^2 \sim \rho^2$ vanishing can compensate.

Combining these constraints, the virial identity is valid for
\begin{equation}
1 - n < \alpha < \alpha_{\max},
\label{eq:alpha_range}
\end{equation}
where $\alpha_{\max} = \infty$ for massive theories and $\alpha_{\max} = 2p + 3 - n$ for power-law decay. The lower bound $\alpha > 1 - n$ ensures core convergence; the upper bound ensures tail convergence. Within this range, each $\alpha$ provides an independent constraint, and the family collectively probes the entire radial structure.

\subsection{Derrick's theorem}

Setting $\alpha = 1$ in Eq.~\eqref{eq:scalar_virial} yields $(2-n)T = nU$, which is Derrick's theorem~\cite{Derrick1964}. For $n \geq 3$ with $T, U > 0$ (positive-definite kinetic energy and non-negative potential bounded below), this constraint cannot be satisfied: the kinetic and potential terms scale in the same direction under uniform dilation, and no equilibrium exists. This explains the absence of stable scalar solitons in three or more spatial dimensions.

However, Derrick's theorem at $\alpha = 1$ weights the profile with $\rho^{n-1}$, which samples all radii with roughly equal importance (up to the exponential or power-law suppression in the tail). Local errors in the core can be compensated by local errors in the intermediate region, yielding an $\alpha = 1$ identity that is satisfied to high precision even when the solution is inaccurate locally.

By contrast, $\alpha < 1$ (and especially negative $\alpha$) shifts the weight toward small $\rho$. The factor $\rho^{\alpha-1}$ in the kinetic integrand diverges as $\rho \to 0$ for $\alpha < 1$, amplifying the contribution from the core. If finite-difference errors are concentrated where the field gradients are steepest (typically near the core), these errors are amplified by negative-$\alpha$ weighting. The result is a virial discrepancy that the standard $\alpha = 1$ test. The $\alpha$-family thus provides independent constraints that probe different radial regions.

\section{Scalar instantons}
\label{sec:scalar}

\subsection{Fubini-Lipatov instanton}

The Fubini-Lipatov instanton provides the simplest analytical verification of the virial identities in a nontrivial setting, and simultaneously illustrates a pathology of the standard Derrick test: in conformally invariant theories, the $\alpha = 1$ relation is satisfied trivially and provides no constraint whatsoever.

This solution arises in four-dimensional Euclidean scalar field theory with action
\begin{equation}
S_E[\phi] = \int d^4x \left[\frac{1}{2}(\partial_\mu\phi)^2 + V(\phi)\right], \qquad V(\phi) = -\frac{\lambda}{4}\phi^4.
\label{eq:FL_action}
\end{equation}
Although this potential is unbounded from below (raising questions of stability that we set aside here), it admits finite-action O(4) symmetric solutions relevant to conformal anomalies and certain cosmological scenarios.

For O(4) symmetric configurations $\phi = \phi(\rho)$ with $\rho = |x|$, the action reduces to
\begin{equation}
S_E = 2\pi^2 \int_0^\infty \mathcal{G}_\rho \, d\rho, \qquad \mathcal{G}_\rho = \rho^3\left[\frac{1}{2}\dot{\phi}^2 + V(\phi)\right],
\label{eq:FL_reduced}
\end{equation}
where $\dot{\phi} = d\phi/d\rho$. The auxiliary quantity~\eqref{eq:C_general} becomes
\begin{equation}
\mathcal{C}_\rho = \rho^3\left[\frac{1}{2}\dot{\phi}^2 - V(\phi)\right].
\label{eq:FL_C}
\end{equation}

The exact instanton solution, discovered independently by Fubini and Lipatov~\cite{Rajaraman1982}, takes the form
\begin{equation}
\phi(\rho) = \frac{\phi_0 R}{\rho^2 + R^2}, \qquad \phi_0 = \sqrt{\frac{8}{\lambda}},
\label{eq:FL_solution}
\end{equation}
where $R > 0$ is an arbitrary scale parameter. The existence of this free parameter reflects the conformal (scale) invariance of the massless $\phi^4$ theory in four dimensions: under the rescaling $\rho \to \lambda\rho$, $\phi \to \lambda^{-1}\phi$, the action remains invariant. Consequently, solutions related by different choices of $R$ are physically equivalent and form a one-parameter family (the ``moduli space'' of instantons).

We now verify that the $\alpha$-family is satisfied for all valid $\alpha$. From the solution~\eqref{eq:FL_solution}, the radial derivative is
\begin{equation}
\dot{\phi} = -\frac{2\phi_0 R\rho}{(\rho^2 + R^2)^2},
\end{equation}
and the kinetic and potential densities evaluate to
\begin{equation}
\frac{1}{2}\dot{\phi}^2 = \frac{2\phi_0^2 R^2 \rho^2}{(\rho^2 + R^2)^4}, \qquad V(\phi) = -\frac{2\phi_0^2 R^4}{(\rho^2 + R^2)^4}.
\end{equation}
Both densities exhibit $\rho^{-6}$ power-law decay at large $\rho$, ensuring convergence of the action integral but limiting the valid $\alpha$ range from above.

The auxiliary quantity $\mathcal{C}_\rho$ defined in Eq.~\eqref{eq:C_general} becomes
\begin{equation}
\mathcal{C}_\rho = \rho^3 \left[\frac{1}{2}\dot{\phi}^2 - V\right] = \frac{2\phi_0^2 R^2 \rho^3 (\rho^2 + R^2)}{(\rho^2 + R^2)^4} = \frac{2\phi_0^2 R^2 \rho^3}{(\rho^2 + R^2)^3}.
\end{equation}
To verify the virial identity, we compute
\begin{equation}
\frac{d}{d\rho}(\rho^\alpha \mathcal{C}_\rho) = \frac{2\phi_0^2 R^2 \rho^{\alpha+2}[(\alpha-3)\rho^2 + (\alpha+3)R^2]}{(\rho^2 + R^2)^4}.
\end{equation}
The virial identity asserts that $\int_0^\infty \frac{d}{d\rho}(\rho^\alpha \mathcal{C}_\rho)\, d\rho = 0$ whenever the boundary terms vanish. Using the Beta function integral
\begin{equation}
\int_0^\infty \frac{\rho^m}{(\rho^2 + R^2)^4}\, d\rho = \frac{R^{m-7}}{2} B\left(\frac{m+1}{2}, \frac{7-m}{2}\right),
\end{equation}
valid for $-1 < m < 7$, we evaluate both terms in the integrand. Setting $m = \alpha + 4$ for the $(\alpha - 3)\rho^2$ term and $m = \alpha + 2$ for the $(\alpha + 3)R^2$ term, and using the Gamma function recurrence $\Gamma(z+1) = z\Gamma(z)$, we find that the Beta functions satisfy
\begin{equation}
B\left(\frac{\alpha+5}{2}, \frac{3-\alpha}{2}\right) = \frac{\alpha+3}{5-\alpha} B\left(\frac{\alpha+3}{2}, \frac{5-\alpha}{2}\right).
\end{equation}
When combined with the prefactors $(\alpha - 3)$ and $(\alpha + 3)R^2/R^2 = (\alpha + 3)$, the two contributions cancel:
\begin{equation}
(\alpha - 3) \cdot \frac{\alpha+3}{5-\alpha} + (\alpha + 3) \cdot \frac{3-\alpha}{5-\alpha} = \frac{\alpha+3}{5-\alpha}[(\alpha - 3) + (3 - \alpha)] = 0.
\end{equation}
This exact cancellation confirms that the virial identity is satisfied for all $\alpha$ in the range $-3 < \alpha < 3$, where the boundary terms vanish and the integrals converge.

The conformal invariance of this system manifests distinctively within our formalism. At $\alpha = 1$, the standard Derrick relation~\eqref{eq:ratio} with $n = 4$ gives $I_T^1/I_V^1 = -2$. However, because both $I_T^\alpha$ and $I_V^\alpha$ scale as $R^{\alpha-1}$ under $R \to \lambda R$, this relation provides no constraint on the instanton size $R$: it is satisfied trivially for any $R$. The $\alpha = 1$ test confirms correctness but provides no information about the profile shape. By contrast, the $\alpha \neq 1$ relations constrain the \emph{distribution} of action density even when the total action is scale-invariant. A numerical solution that deviates from the exact profile would violate some $\alpha \neq 1$ identity, even if $\alpha = 1$ is satisfied.

The Euclidean action can be computed directly:
\begin{equation}
S_E = 2\pi^2 \int_0^\infty \left[\frac{1}{2}\dot{\phi}^2 + V(\phi)\right] \rho^3\, d\rho = \frac{8\pi^2}{3\lambda},
\end{equation}
independent of $R$ as required by scale invariance.

The virial identity requires $[\rho^\alpha \mathcal{C}_\rho]_0^\infty = 0$. From the solution~\eqref{eq:FL_solution}:

\noindent\textit{At the origin} ($\rho \to 0$): The field approaches a constant $\phi(0) = \phi_0/R$, and the derivative vanishes as
\begin{equation}
\dot{\phi} = -\frac{2\phi_0 R \rho}{(\rho^2 + R^2)^2} \sim -\frac{2\phi_0 \rho}{R^3} \quad \Rightarrow \quad \dot{\phi}^2 \sim \rho^2.
\end{equation}
The potential $V(\phi) \to V(\phi_0/R) = -\lambda\phi_0^4/(4R^4)$ remains finite. Thus $\mathcal{C}_\rho = \rho^3[\frac{1}{2}\dot{\phi}^2 - V] \sim \rho^3$, and
\begin{equation}
\rho^\alpha \mathcal{C}_\rho \sim \rho^{\alpha+3} \to 0 \quad \text{for } \alpha > -3.
\end{equation}

\noindent\textit{At infinity} ($\rho \to \infty$): The field decays as $\phi \sim \phi_0 R/\rho^2$, giving
\begin{equation}
\dot{\phi} \sim -\frac{2\phi_0 R}{\rho^3}, \quad \dot{\phi}^2 \sim \rho^{-6}, \quad V \sim -\frac{\lambda\phi_0^4 R^4}{4\rho^8} \sim \rho^{-8}.
\end{equation}
The kinetic term dominates, so $\mathcal{C}_\rho \sim \rho^3 \cdot \rho^{-6} = \rho^{-3}$, and
\begin{equation}
\rho^\alpha \mathcal{C}_\rho \sim \rho^{\alpha-3} \to 0 \quad \text{for } \alpha < 3.
\end{equation}

The identities are valid for $-3 < \alpha < 3$. The finite upper bound reflects the power-law decay characteristic of conformally invariant theories.

\subsection{Vacuum decay}

The decay of a metastable vacuum proceeds via quantum tunneling, described semiclassically by an O(4) symmetric bounce solution~\cite{Coleman1977}. The tunneling rate per unit volume takes the form $\Gamma/V \sim A e^{-B}$, where $B$ is the Euclidean action of the bounce and $A$ is a prefactor arising from fluctuations around the classical solution. Because $B$ enters the exponential, even modest fractional errors in the bounce action translate into order-of-magnitude uncertainties in the decay rate.

The bounce configuration interpolates between the false vacuum at spatial infinity and a region near the true vacuum at the origin, representing the nucleation of a critical bubble. The Euclidean action is
\begin{equation}
S_E[\phi] = \int d^4x \left[\frac{1}{2}(\partial_\mu\phi)^2 + V(\phi)\right],
\label{eq:bounce_action}
\end{equation}
with the quartic potential featuring explicit symmetry breaking:
\begin{equation}
V(\phi) = \frac{\lambda}{8}(\phi^2 - a^2)^2 - \frac{\epsilon}{2}(\phi + a).
\label{eq:bounce_potential}
\end{equation}
For small $\epsilon > 0$, this potential has a metastable false vacuum near $\phi_{\rm fv} \approx -a$ and a deeper true vacuum near $\phi_{\rm tv} \approx +a$.

For O(4) symmetric configurations $\phi = \phi(\rho)$, the reduced functional and auxiliary quantity are
\begin{equation}
\mathcal{G}_\rho = \rho^3\left[\frac{1}{2}\dot{\phi}^2 + V(\phi)\right], \qquad \mathcal{C}_\rho = \rho^3\left[\frac{1}{2}\dot{\phi}^2 - V(\phi)\right],
\label{eq:bounce_GC}
\end{equation}
with $V$ shifted so that $V(\phi_{\rm fv}) = 0$. The bounce satisfies $\dot{\phi}(0) = 0$ (regularity at the origin) and $\phi(\infty) = \phi_{\rm fv}$ (approach to the false vacuum).

For $n = 4$, the scalar virial relation~\eqref{eq:scalar_virial} gives
\begin{equation}
(\alpha - 3) I_T^\alpha = (\alpha + 3) I_V^\alpha,
\label{eq:bounce_virial}
\end{equation}
where the weighted integrals are
\begin{equation}
I_T^\alpha = \int_0^\infty \frac{1}{2}\dot{\phi}^2 \rho^{\alpha+2}\, d\rho, \qquad I_V^\alpha = \int_0^\infty V(\phi) \rho^{\alpha+2}\, d\rho,
\end{equation}
and $V$ is shifted so that $V(\phi_{\rm fv}) = 0$. The normalization $V(\phi_\mathrm{fv})=0$ is
required by the vanishing of the boundary term. On a finite interval $[0,R]$,
the identity acquires a boundary contribution
$R^{\alpha+n-1}(\frac{1}{2}\dot\phi(R)^2 - V(\phi(R)))$; for solutions with
infinite support this must vanish as $R\to\infty$, while for compactons with
$\dot\phi(R)=0$ it vanishes at finite~$R$. Both cases require
$V(\phi_\mathrm{fv})=0$. The ratio $I_T^\alpha/I_V^\alpha = (\alpha + 3)/(\alpha - 3)$ takes specific values at each $\alpha$: $-1$ at $\alpha = 0$, $-2$ at $\alpha = 1$, and $-5$ at $\alpha = 2$.

The virial relation at $\alpha = 1$ yields an important result. Setting $\alpha = 1$:
\begin{equation}
-2 I_T^1 = 4 I_V^1, \quad \text{i.e.,} \quad \int_0^\infty \dot{\phi}^2 \rho^3\, d\rho = -4 \int_0^\infty V(\phi) \rho^3\, d\rho.
\end{equation}
The bounce action is
\begin{equation}
B = 2\pi^2 \int_0^\infty \left[\frac{1}{2}\dot{\phi}^2 + V(\phi)\right] \rho^3\, d\rho.
\end{equation}
Using the virial relation to eliminate the potential integral:
\begin{equation}
B = 2\pi^2 \int_0^\infty \left[\frac{1}{2}\dot{\phi}^2 - \frac{1}{4}\dot{\phi}^2\right] \rho^3\, d\rho = \frac{\pi^2}{2} \int_0^\infty \dot{\phi}^2 \rho^3\, d\rho.
\end{equation}
Since $\dot{\phi}^2 \geq 0$ and the bounce is nontrivial, we conclude that $B > 0$. This positivity ensures exponential suppression of the decay rate $\Gamma \sim e^{-B}$, a result first established by Coleman~\cite{Coleman1977} and Weinberg~\cite{Weinberg1996}.

Treating the radial coordinate $\rho$ as ``time,'' the bounce equation
\begin{equation}
\ddot{\phi} + \frac{3}{\rho}\dot{\phi} = \frac{dV}{d\phi}
\end{equation}
describes a particle moving in the inverted potential $-V(\phi)$ subject to a time-dependent friction force $-(3/\rho)\dot{\phi}$. The friction is strongest at small ``times'' (small $\rho$) and diminishes as $\rho$ increases. The particle must start at rest at some initial ``height'' $\phi(0)$ on the ``hill'' corresponding to the inverted true vacuum and coast to rest exactly at the inverted false vacuum as $\rho \to \infty$. The friction dissipates the ``energy'' acquired during the descent, allowing the particle to come to rest.

This analogy connects naturally to our formalism at the boundary of the validity range. At $\alpha = -3$, the weighted quantity
\begin{equation}
\mathcal{C}_\rho\big|_{\alpha=-3} = \frac{1}{2}\dot{\phi}^2 - V(\phi) \equiv E_{\rm mech}
\end{equation}
coincides with the mechanical energy in Coleman's analogy ~\cite{Coleman1977}. The evolution equation~\eqref{eq:C_evolution_general} at $\alpha = -3$ becomes
\begin{equation}
\frac{d}{d\rho}\left[\frac{1}{2}\dot{\phi}^2 - V\right] = -\frac{3}{\rho}\dot{\phi}^2,
\end{equation}
which states that mechanical energy is dissipated by friction, the right-hand side is manifestly negative for $\rho > 0$.

The integral form, accounting for the non-vanishing boundary term, reads
\begin{equation}
-3 \int_0^\infty \dot{\phi}^2 \rho^{-1}\, d\rho = \left[\frac{1}{2}\dot{\phi}^2 - V(\phi)\right]\bigg|_0^\infty.
\end{equation}
While the simplified virial identity itself fails at $\alpha = -3$ (the boundary term is nonzero), this limiting case clarifies the role of the friction term.

\begin{figure}[t]
\centering
\includegraphics[width=0.9\columnwidth]{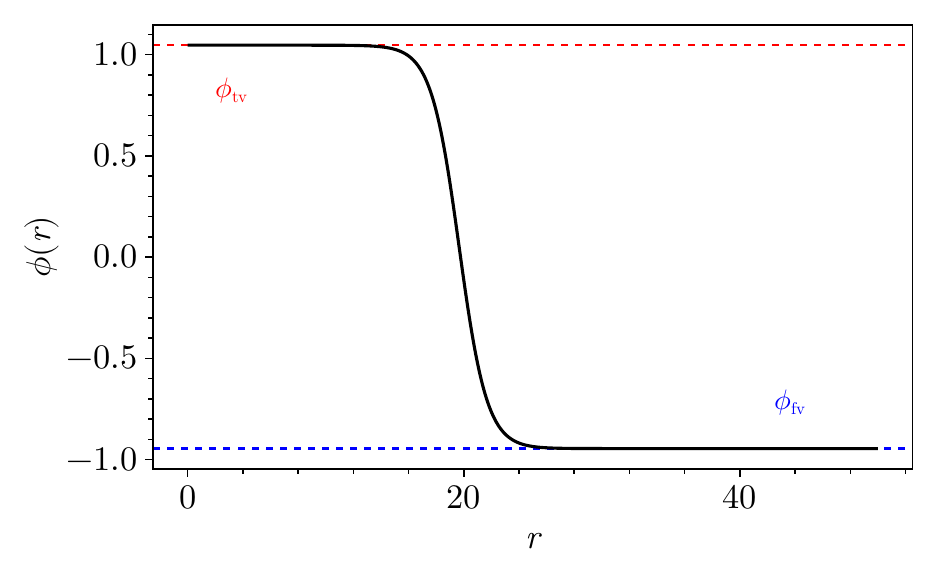}
\caption{Bounce profile $\phi(\rho)$ for the potential~\eqref{eq:bounce_potential} with $\lambda = 1$, $a = 1$, and $\epsilon = 0.1$. The field interpolates from the true vacuum $\phi_{\rm tv} \approx 1.00$ at the origin to the false vacuum $\phi_{\rm fv} \approx -0.95$ at large $\rho$. In the thin-wall regime, the transition occurs at large radius, making the asymptotic approach to the false vacuum particularly important.}
\label{fig:bounce}
\end{figure}

We computed the bounce numerically for the potential (3.18) with
$\lambda=1$, $a=1$, and $\epsilon=0.1$
using the \texttt{AnyBubble} package~\cite{Masoumi:2016wot}. The resulting profile is shown in Figure~\ref{fig:bounce}. The bounce action for these parameters is $B \approx 24725.2$. Table~\ref{tab:bounce} presents the numerical verification of the virial relations across a range of $\alpha$ values.

The pattern of errors across $\alpha$ shows the radial distribution of numerical inaccuracies. The error grows monotonically with $\alpha$, from $0.001\%$ at $\alpha = -2$ to $0.03\%$ at $\alpha = 2$. Since large positive $\alpha$ emphasizes the tail while negative $\alpha$ emphasizes the core, this pattern indicates that the numerical errors are concentrated in the asymptotic region rather than at the origin. This makes sense: the bounce approaches the false vacuum exponentially as $\rho \to \infty$, and the numerical solution must be truncated at some finite radius $\rho_{\rm max}$. In the thin-wall regime (small $\epsilon$), the bubble wall sits at large radius, making accurate resolution of the tail particularly demanding.

\begin{table}[h]
\centering
\caption{Numerical verification of virial relations for the Coleman bounce with parameters $\lambda = 1$, $a = 1$, $\epsilon = 0.1$. The LHS and RHS denote $(\alpha - 3)I_T^\alpha$ and $(\alpha + 3)I_V^\alpha$ respectively. Digits where the two sides differ are shown in \textcolor{red}{red}. The error grows monotonically with $\alpha$, indicating that numerical inaccuracies are concentrated in the tail region.}
\label{tab:bounce}
\begin{tabular}{cccc}
\toprule
$\alpha$ & LHS & RHS & Error (\%) \\
\midrule
$-2.0$ & $-1.630910$ & $-1.6309\textcolor{red}{30}$ & $0.0011$ \\
$-1.0$ & $-25.65572$ & $-25.655\textcolor{red}{64}$ & $0.0029$ \\
$0.0$ & $-379.659$ & $-379.6\textcolor{red}{83}$ & $0.0059$ \\
$1.0$ & $-5010.98$ & $-501\textcolor{red}{1.60}$ & $0.012$ \\
$2.0$ & $-49770.3$ & $-497\textcolor{red}{86.05}$ & $0.032$ \\
\bottomrule
\end{tabular}
\end{table}

The bounce interpolates between vacua rather than decaying to zero, changing the asymptotic structure.

\noindent\textit{At the origin} ($\rho \to 0$): Regularity requires $\dot{\phi}(0) = 0$, so
\begin{equation}
\phi(\rho) = \phi_0 + \tfrac{1}{2}\phi''(0)\rho^2 + O(\rho^4), \quad \dot{\phi} \sim \phi''(0)\rho, \quad \dot{\phi}^2 \sim \rho^2.
\end{equation}
With $V(\phi_0)$ finite, $\mathcal{C}_\rho = \rho^3[\frac{1}{2}\dot{\phi}^2 - V] \sim \rho^3$, and
\begin{equation}
\rho^\alpha \mathcal{C}_\rho \sim \rho^{\alpha+3} \to 0 \quad \text{for } \alpha > -3.
\end{equation}

\noindent\textit{At infinity} ($\rho \to \infty$): The field approaches the false vacuum exponentially,
\begin{equation}
\phi - \phi_{\rm fv} \sim e^{-m\rho}, \quad \dot{\phi} \sim -m\, e^{-m\rho}, \quad m^2 = V''(\phi_{\rm fv}).
\end{equation}
Both $\dot{\phi}^2$ and $V - V(\phi_{\rm fv})$ decay as $e^{-2m\rho}$, so $\mathcal{C}_\rho \sim \rho^3 e^{-2m\rho}$. Exponential decay defeats any polynomial weight:
\begin{equation}
\rho^\alpha \mathcal{C}_\rho \sim \rho^{\alpha+3} e^{-2m\rho} \to 0 \quad \text{for all finite } \alpha.
\end{equation}

The virial relations are valid for $\alpha > -3$, with no upper bound. The absence of an upper bound reflects the massive nature of the theory near the false vacuum.

\section{Gauge fields}
\label{sec:gauge}

\subsection{BPS monopole}

The 't Hooft-Polyakov monopole~\cite{tHooft1974,Polyakov1974} offers a stringent test of the
virial formalism in gauge theories. This solution arises in
the Georgi--Glashow model~\cite{Georgi:1972cj}, i.e.,SU(2)
Yang-Mills-Higgs theory with Lagrangian
\begin{equation}
\mathcal{L} = -\frac{1}{4}F^a_{\mu\nu}F^{a\mu\nu} + \frac{1}{2}(D_\mu\Phi^a)(D^\mu\Phi^a) - \frac{\lambda}{4}(|\Phi|^2 - v^2)^2,
\label{eq:monopole_lagrangian}
\end{equation}
where $\Phi^a$ is a triplet Higgs field breaking SU(2) to U(1). In the BPS limit~\cite{Bogomolny1976} $\lambda \to 0$, the monopole saturates a topological energy bound and admits an exact analytical solution.

The spherically symmetric hedgehog ansatz is
\begin{equation}
\Phi^a = v H(\rho) \frac{x^a}{\rho}, \qquad A^a_i = \epsilon_{aij}\frac{x^j}{\rho^2}[1 - K(\rho)],
\label{eq:monopole_ansatz}
\end{equation}
where $\rho$ is the dimensionless radial coordinate (in units of $(gv)^{-1}$). Substituting into the energy functional yields the reduced form
\begin{equation}
E = \frac{4\pi v}{g} \int_0^\infty \mathcal{G}_\rho \, d\rho, \qquad \mathcal{G}_\rho = (\dot{K})^2 + \frac{(K^2-1)^2}{\rho^2} + (\rho \dot{H})^2 + K^2 H^2,
\label{eq:monopole_energy}
\end{equation}
where $K(\rho)$ is the gauge profile and $H(\rho)$ is the Higgs profile.

The auxiliary quantity~\eqref{eq:C_general} for this two-field system is
\begin{equation}
\mathcal{C}_\rho = \dot{K}\frac{\partial \mathcal{G}_\rho}{\partial \dot{K}} + \dot{H}\frac{\partial \mathcal{G}_\rho}{\partial \dot{H}} - \mathcal{G}_\rho = (\dot{K})^2 + \rho^2(\dot{H})^2 - \frac{(K^2-1)^2}{\rho^2} - K^2 H^2.
\label{eq:monopole_C}
\end{equation}
The explicit $\rho$-dependence in $\mathcal{G}_\rho$ arises from the geometric weights $\rho^{-2}$ on the magnetic term and $\rho^2$ on the Higgs kinetic term, reflecting the angular structure of the hedgehog configuration.

The BPS monopole satisfies first-order Bogomolny equations:
\begin{equation}
\dot{K} = -KH, \qquad \dot{H} = -\frac{K^2 - 1}{\rho^2}.
\label{eq:BPS_equations}
\end{equation}
These admit the exact solution
\begin{equation}
K(\rho) = \frac{\rho}{\sinh\rho}, \qquad H(\rho) = \coth\rho - \frac{1}{\rho}.
\label{eq:BPS_solution}
\end{equation}

The Bogomolny equations imply \emph{pointwise} identities between kinetic and potential densities:
\begin{equation}
(\dot{K})^2 = K^2 H^2, \qquad (\rho \dot{H})^2 = \frac{(K^2-1)^2}{\rho^2}.
\end{equation}
When integrated with any weight $\rho^{\alpha-1}$, these become
\begin{equation}
I_{\dot{K}}^\alpha = I_{KH}^\alpha, \qquad I_{\dot{H}}^\alpha = I_{K^2}^\alpha,
\label{eq:BPS_identities}
\end{equation}
where $I_{\dot{K}}^\alpha = \int_0^\infty (\dot{K})^2 \rho^{\alpha-1}\, d\rho$, etc. This is the hallmark of BPS configurations: the first-order BPS equations are equivalent to the condition of vanishing pressure, from which the trivial satisfaction of all integral relations follows immediately.

The general virial identity~\eqref{eq:general_virial} for this system reads
\begin{equation}
\alpha I_{\dot{K}}^\alpha + (\alpha - 2) I_{\dot{H}}^\alpha = (\alpha - 2) I_{K^2}^\alpha + \alpha I_{KH}^\alpha.
\label{eq:monopole_virial}
\end{equation}
Using the BPS identities~\eqref{eq:BPS_identities}, both sides reduce to $\alpha I_{\dot{K}}^\alpha + (\alpha - 2) I_{\dot{H}}^\alpha$, so the identity is satisfied tautologically for all valid $\alpha$.

At $\alpha = 2$, the identity~\eqref{eq:monopole_virial} reduces to $I_{\dot{K}}^2 = I_{KH}^2$: the gauge kinetic moment equals the gauge-Higgs coupling moment, with no contribution from the Higgs sector. This special case probes the intermediate region where both fields vary significantly. At $\alpha = 0$, we obtain a constraint dominated by the core, where the centrifugal barrier is strongest.

For the BPS monopole, the Bogomolny equations imply $\mathcal{C}_\rho = 0$ pointwise, so the boundary term vanishes identically.

\noindent\textit{At the origin} ($\rho \to 0$): The regular solution behaves as
\begin{equation}
K = 1 - \frac{\rho^2}{6} + O(\rho^4), \quad H = \frac{\rho}{3} + O(\rho^3).
\end{equation}
Thus $\dot{K} \sim -\rho/3$, $\dot{H} \sim 1/3$, and each term in $\mathcal{G}_\rho$ scales as $\rho^2$:
\begin{equation}
\dot{K}^2 \sim \rho^2, \quad \rho^2\dot{H}^2 \sim \rho^2, \quad \frac{(K^2-1)^2}{\rho^2} \sim \rho^2, \quad K^2 H^2 \sim \rho^2.
\end{equation}
The integral $I_{\dot{K}}^\alpha = \int \dot{K}^2 \rho^{\alpha-1} d\rho$ has integrand $\sim \rho^{\alpha+1}$, converging for $\alpha > -2$.

\noindent\textit{At infinity} ($\rho \to \infty$): The unbroken $U(1)$ symmetry produces a massless photon whose Coulomb field gives power-law tails in both the gauge and Higgs sectors. From the exact solution~\eqref{eq:BPS_solution}, the gauge profile decays exponentially, $K \sim 2\rho\,e^{-\rho}$, while the Higgs profile approaches its VEV algebraically, $H - 1 \sim -1/\rho$. Of the four integrals in~\eqref{eq:monopole_virial}, $I_{\dot{H}}^\alpha$ and $I_{K^2}^\alpha$ have integrands falling as $\rho^{\alpha-3}$ and diverge individually for $\alpha \geq 2$, while $I_{\dot{K}}^\alpha$ and $I_{KH}^\alpha$ decay exponentially. The Bogomolny equations enforce exact pairwise cancellations: $I_{\dot{K}}^\alpha = I_{KH}^\alpha$ and $I_{\dot{H}}^\alpha = I_{K^2}^\alpha$, so the virial identity reduces to $0 = 0$ for all~$\alpha$. No upper bound on~$\alpha$ arises from infinity.

Thus the relations hold for $\alpha > -2$. The pointwise BPS condition $\mathcal{C}_\rho = 0$ ensures the virial identity is satisfied throughout this range.

\subsection{BPST instanton}

The BPST instanton~\cite{BPST1975} in four-dimensional pure Yang-Mills theory
provides a second example of BPS triviality, where the self-duality
equation $F = \ast F$ guarantees that the virial identities are satisfied
for all valid~$\alpha$. Additionally, conformal invariance renders the
$\alpha = 1$ identity degenerate. The Euclidean action is
\begin{equation}
S_E = \frac{1}{2g^2}\int d^4x \, \text{Tr}(F_{\mu\nu}F^{\mu\nu}).
\label{eq:YM_full_action}
\end{equation}

The O(4) symmetric ansatz for SU(2) gauge fields,
\begin{equation}
A_\mu^a = \frac{2}{g}\bar{\eta}^a_{\mu\nu}\frac{x^\nu}{\rho^2}w(\rho),
\label{eq:BPST_ansatz}
\end{equation}
where $\bar{\eta}^a_{\mu\nu}$ is the 't~Hooft symbol, reduces the action to
\begin{equation}
S_E = \frac{24\pi^2}{g^2} \int_0^\infty \mathcal{G}_\rho \, d\rho, \qquad \mathcal{G}_\rho = \frac{1}{2}\rho \dot{w}^2 + \frac{2}{\rho} w^2(1-w)^2,
\label{eq:YM_action}
\end{equation}
where the profile function $w(\rho)$ satisfies $w(0) = 0$ and $w(\infty) = 1$.

The auxiliary quantity is
\begin{equation}
\mathcal{C}_\rho = \frac{1}{2}\rho\dot{w}^2 - \frac{2}{\rho}w^2(1-w)^2.
\label{eq:BPST_C}
\end{equation}
The kinetic term carries weight $\rho$ from the four-dimensional measure, while the potential term carries $\rho^{-1}$, acting as a centrifugal-like barrier that concentrates action density near the instanton center.

The BPST solution is
\begin{equation}
w(\rho) = \frac{\rho^2}{\rho^2 + \rho_0^2},
\label{eq:BPST_solution}
\end{equation}
where $\rho_0 > 0$ is the instanton size, an arbitrary parameter reflecting conformal invariance.

The self-duality equation $F = \ast F$ reduces in the radial
problem to
\begin{equation}
\rho\dot{w} = 2w(1-w).
\label{eq:BPST_BPS}
\end{equation}
Squaring both sides gives $\rho^2\dot{w}^2 = 4w^2(1-w)^2$, i.e.,
the kinetic and potential densities in~\eqref{eq:YM_action} are related
pointwise:
\begin{equation}
\frac{1}{2}\rho\dot{w}^2 = \frac{2}{\rho}w^2(1-w)^2.
\end{equation}
This implies $\mathcal{C}_\rho = 0$ exactly, just as for the BPS
monopole. The virial identities are therefore trivially satisfied for
all valid~$\alpha$.

Applying the general virial formula~\eqref{eq:general_virial}, both kinetic and potential coefficients are proportional to $(\alpha - 1)$:
\begin{equation}
(\alpha - 1) \int_0^\infty \dot{w}^2 \rho^\alpha\, d\rho = 4(\alpha - 1) \int_0^\infty w^2(1-w)^2 \rho^{\alpha-2}\, d\rho.
\label{eq:YM_virial_pre}
\end{equation}
At $\alpha = 1$, both sides vanish identically: the standard Derrick scaling argument provides no constraint because the action is scale-invariant. This degeneracy is specific to conformally invariant theories and is independent of the BPS condition.

For $\alpha \neq 1$, dividing by $(\alpha - 1)$ yields
\begin{equation}
\int_0^\infty \dot{w}^2 \rho^\alpha\, d\rho = 4 \int_0^\infty w^2(1-w)^2 \rho^{\alpha-2}\, d\rho,
\label{eq:YM_virial}
\end{equation}
which is the self-duality equation squared and integrated with
weight $\rho^{\alpha-2}$. Both integrals scale as $\rho_0^{\alpha-1}$
under rescaling, so the identity is satisfied for any $\rho_0$ as required
by conformal invariance.

Analytical verification confirms~\eqref{eq:YM_virial}: using the Beta function integral, both sides evaluate to $\frac{1}{2}\rho_0^{\alpha-1} B((\alpha+3)/2, (5-\alpha)/2)$. The boundary terms vanish and the integrals converge for $-3 < \alpha < 5$, and the identity holds throughout this range.

From the BPST solution $w = \rho^2/(\rho^2 + \rho_0^2)$:

\noindent\textit{At the origin} ($\rho \to 0$): The profile vanishes as $w \sim \rho^2/\rho_0^2$, giving
\begin{equation}
\dot{w} = \frac{2\rho\rho_0^2}{(\rho^2 + \rho_0^2)^2} \sim \frac{2\rho}{\rho_0^2}, \quad \dot{w}^2 \sim \rho^2.
\end{equation}
The potential term: $w^2(1-w)^2 \sim \rho^4/\rho_0^4$. Thus
\begin{equation}
\mathcal{C}_\rho = \tfrac{1}{2}\rho\dot{w}^2 - \tfrac{2}{\rho}w^2(1-w)^2 \sim \rho^3 - \rho^3 \sim \rho^3,
\end{equation}
and $\rho^\alpha \mathcal{C}_\rho \sim \rho^{\alpha+3} \to 0$ for $\alpha > -3$.

\noindent\textit{At infinity} ($\rho \to \infty$): The profile approaches unity as $1 - w \sim \rho_0^2/\rho^2$, giving
\begin{equation}
\dot{w} \sim \frac{2\rho_0^2}{\rho^3}, \quad \dot{w}^2 \sim \rho^{-6}, \quad w^2(1-w)^2 \sim \frac{\rho_0^4}{\rho^4}.
\end{equation}
Both terms in $\mathcal{C}_\rho$ scale as $\rho^{-5}$:
\begin{equation}
\rho\dot{w}^2 \sim \rho^{-5}, \quad \frac{w^2(1-w)^2}{\rho} \sim \rho^{-5},
\end{equation}
so $\mathcal{C}_\rho \sim \rho^{-5}$ and $\rho^\alpha \mathcal{C}_\rho \sim \rho^{\alpha-5} \to 0$ for $\alpha < 5$.

In this case the relations hold for $-3 < \alpha < 5$. Like the Fubini-Lipatov instanton, the finite upper bound reflects power-law decay in this conformally invariant theory.

\subsection{Nielsen-Olesen vortices}
\label{sec:vortex}

The Nielsen-Olesen vortex~\cite{NielsenOlesen1973} is the fundamental topological defect in the Abelian Higgs model with Lagrangian
\begin{equation}
\mathcal{L} = -\frac{1}{4}F_{\mu\nu}F^{\mu\nu} + |D_\mu\phi|^2 - \frac{\lambda}{4}(|\phi|^2 - v^2)^2,
\label{eq:vortex_lagrangian}
\end{equation}
where $D_\mu = \partial_\mu - ieA_\mu$ and $\phi$ is a complex scalar.

For a unit-winding vortex with cylindrical symmetry, the ansatz
\begin{equation}
\phi = vf(\rho)e^{i\theta}, \qquad A_\theta = \frac{1}{e\rho}[1 - a(\rho)],
\label{eq:vortex_ansatz}
\end{equation}
reduces the static energy to
\begin{equation}
E = 2\pi v^2 \int_0^\infty \mathcal{G}_\rho \, d\rho, \qquad \mathcal{G}_\rho = \frac{1}{2\rho}\dot{a}^2 + \rho\dot{f}^2 + \frac{1}{\rho}f^2 a^2 + \frac{\beta \rho}{4}(f^2 - 1)^2,
\label{eq:vortex_energy}
\end{equation}
where $a(\rho)$ is the gauge profile, $f(\rho)$ is the scalar profile, $\beta = \lambda/e^2$ determines the vortex type, and dots denote radial derivatives.

The auxiliary quantity for this two-field system is
\begin{equation}
\mathcal{C}_\rho = \frac{1}{\rho}\dot{a}^2 + \rho\dot{f}^2 - \frac{1}{\rho}f^2a^2 - \frac{\beta\rho}{4}(f^2-1)^2.
\label{eq:vortex_C}
\end{equation}
The gauge kinetic term $\dot{a}^2/(2\rho)$ and covariant coupling $f^2a^2/\rho$ share the weight $\rho^{-1}$, reflecting the magnetic field structure $B \propto \dot{a}/\rho$. The scalar kinetic term $\rho\dot{f}^2$ and Higgs potential carry the standard two-dimensional weight $\rho$.

Defining the $\alpha$-weighted integrals
\begin{align}
I_a^\alpha &= \int_0^\infty \dot{a}^2 \rho^{\alpha-2}\, d\rho, & I_f^\alpha &= \int_0^\infty \dot{f}^2 \rho^\alpha\, d\rho, \\
I_{fa}^\alpha &= \int_0^\infty f^2 a^2 \rho^{\alpha-2}\, d\rho, & I_H^\alpha &= \int_0^\infty (f^2 - 1)^2 \rho^\alpha\, d\rho,
\end{align}
the virial relation takes the form
\begin{equation}
\frac{\alpha + 1}{2} I_a^\alpha + (\alpha - 1) I_f^\alpha = (\alpha - 1) I_{fa}^\alpha + \frac{\beta(\alpha + 1)}{4} I_H^\alpha.
\label{eq:vortex_virial}
\end{equation}

At $\alpha = 1$, the terms proportional to $(\alpha - 1)$ vanish, leaving
\begin{equation}
I_a^1 = \frac{\beta}{2} I_H^1,
\end{equation}
which balances magnetic flux energy against Higgs potential energy, the analogue of Derrick's relation for this system. At $\alpha = -1$, the terms proportional to $(\alpha + 1)$ vanish, leaving
\begin{equation}
I_f^{-1} = I_{fa}^{-1},
\end{equation}
which balances scalar kinetic energy against covariant coupling energy in the core region.

\begin{figure}[tbh]
\centering
\includegraphics[width=0.8\columnwidth]{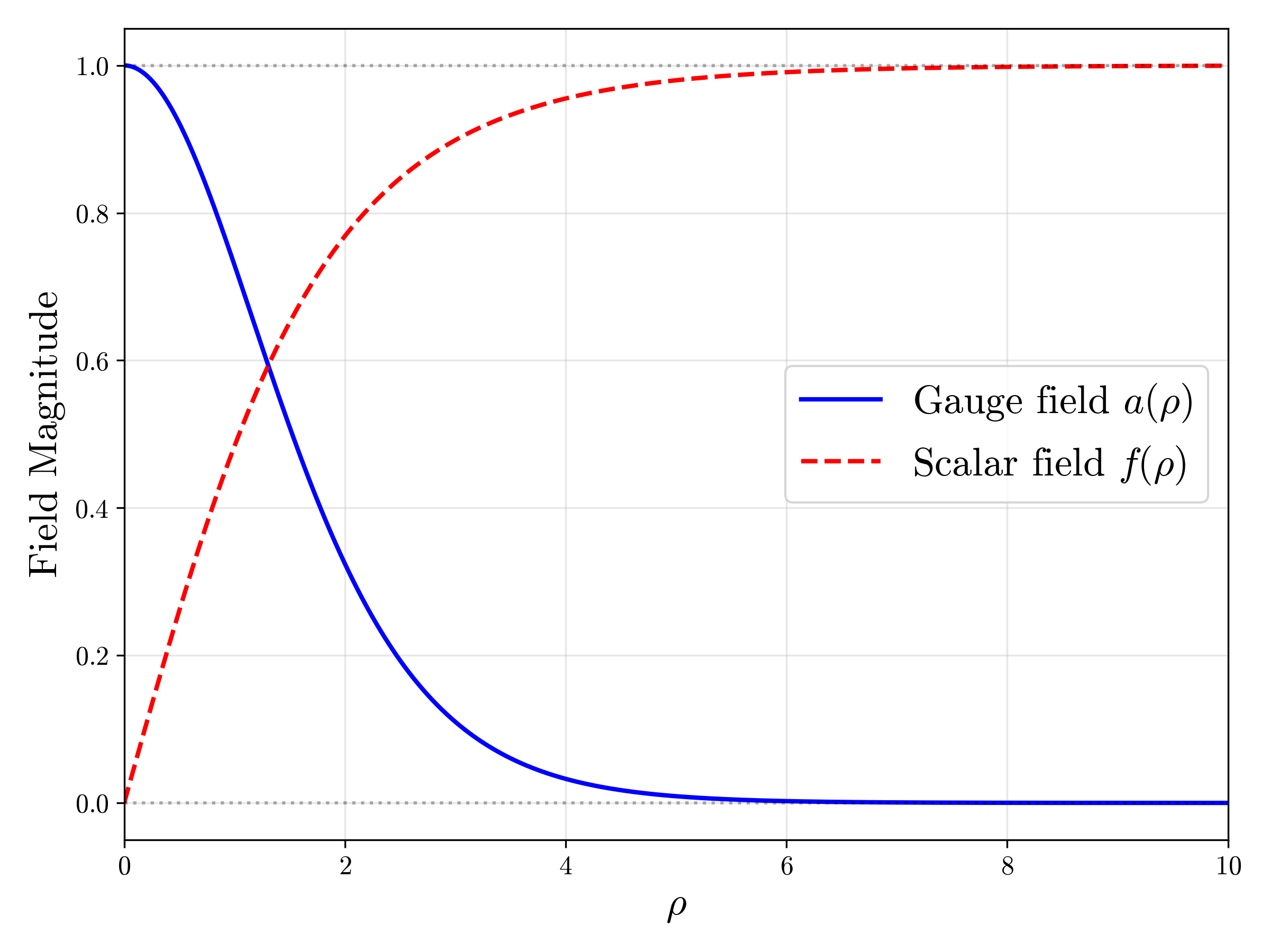}
\caption{Radial profiles of the gauge field $a(\rho)$ (decreasing from 1) and scalar field $f(\rho)$ (increasing from 0) for the Nielsen-Olesen vortex at coupling $\beta = 0.5$ (Type-I regime). The characteristic scales of the two profiles differ: the gauge field penetration depth exceeds the scalar coherence length for $\beta < 1$. The steepest gradients occur in the core region near $\rho = 0$.}
\label{fig:vortex}
\end{figure}

We solved the vortex equations numerically for $\beta = 0.5$ (Type-I regime) using a standard finite-difference scheme on a logarithmic grid. The resulting profiles are shown in Figure~\ref{fig:vortex}. Table~\ref{tab:vortex} presents the numerical verification of Eq.~\eqref{eq:vortex_virial} across a range of $\alpha$ values.

The results illustrate how different $\alpha$ values probe different radial regions. At $\alpha = 1$, the agreement is excellent: the relative error is below $0.001\%$. The standard Derrick test would indicate an accurate solution. However, as $\alpha$ decreases toward negative values, the error grows dramatically, reaching $5.7\%$ at $\alpha = -0.5$.

Negative $\alpha$ emphasizes the core: the weighting $\rho^{\alpha-2}$ in the gauge integrals diverges as $\rho \to 0$ for $\alpha < 2$, amplifying the contribution from the innermost region. Finite-difference errors are typically largest where field gradients are steepest, precisely the core region where $f$ rises from zero and $a$ falls from unity. The $\alpha = -0.5$ identity weights these errors heavily, exposing inaccuracies that the $\alpha = 1$ identity averages away.

Conversely, at large $\alpha$ (e.g., $\alpha = 10$), errors grow to about $0.02\%$ because the weighting $\rho^\alpha$ amplifies numerical uncertainties in the asymptotic tail, the region where the grid becomes coarse and the fields approach their boundary values.

The pattern of errors is characteristic of a solution that is well-converged in the bulk but has slightly larger relative errors in the core and tail regions. The standard Derrick test cannot provide this information: a single number ($0.0005\%$ agreement) masks significant local inaccuracies.

\begin{table}[h]
\centering
\caption{Numerical verification of the vortex virial identity~\eqref{eq:vortex_virial} for $\beta = 0.5$. The LHS and RHS denote the left and right sides of Eq.~\eqref{eq:vortex_virial}. Digits where the two sides differ are shown in \textcolor{red}{red}. The $5.7\%$ error at $\alpha = -0.5$ exposes core-region inaccuracies invisible to the standard Derrick test ($\alpha = 1$), where the error is only $0.0005\%$.}
\label{tab:vortex}
\begin{tabular}{cccc}
\toprule
$\alpha$ & LHS & RHS & Error (\%) \\
\midrule
$-0.5$ & $-0.69034782$ & $-0.6\textcolor{red}{5088331}$ & $5.72$ \\
$0$ & $-0.12322003$ & $-0.12\textcolor{red}{197205}$ & $1.01$ \\
$1$ & $0.29078687$ & $0.29078\textcolor{red}{831}$ & $0.00049$ \\
$3$ & $2.41022217$ & $2.410222\textcolor{red}{28}$ & $0.0000044$ \\
$6$ & $120.54782101$ & $120.547\textcolor{red}{65125}$ & $0.00014$ \\
$8$ & $3375.35429541$ & $3375.\textcolor{red}{28663642}$ & $0.0020$ \\
$10$ & $151493.14599728$ & $1514\textcolor{red}{66.12688095}$ & $0.018$ \\
\bottomrule
\end{tabular}
\end{table}

The vortex profiles satisfy $a(0) = 1$, $f(0) = 0$ (regularity with unit winding) and $a(\infty) = 0$, $f(\infty) = 1$ (vacuum at infinity)

\noindent\textit{At the origin} ($\rho \to 0$): Regularity requires
\begin{equation}
f \sim c_f \rho, \quad a \sim 1 - c_a \rho^2, \quad \Rightarrow \quad \dot{f} \sim c_f, \quad \dot{a} \sim -2c_a \rho.
\end{equation}
Each term in $\mathcal{C}_\rho$ scales linearly in $\rho$:
\begin{equation}
\frac{\dot{a}^2}{\rho} \sim \rho, \quad \rho\dot{f}^2 \sim \rho, \quad \frac{f^2 a^2}{\rho} \sim \rho, \quad \rho(f^2-1)^2 \sim \rho.
\end{equation}
Thus $\mathcal{C}_\rho \sim \rho$ and $\rho^\alpha \mathcal{C}_\rho \sim \rho^{\alpha+1} \to 0$ for $\alpha > -1$.

\noindent\textit{At infinity} ($\rho \to \infty$): Both fields approach their vacuum values exponentially,
\begin{equation}
a \sim e^{-m_A \rho}, \quad f - 1 \sim e^{-m_H \rho},
\end{equation}
where $m_A$ and $m_H$ are the gauge and Higgs masses. All terms in $\mathcal{C}_\rho$ decay exponentially, ensuring convergence for all finite $\alpha$.

Then the integral relations are valid for $\alpha>-1$.

\subsection{Electroweak sphaleron}
The electroweak sphaleron illustrates the virial formalism in a system where the Higgs mechanism explicitly breaks scale invariance, introducing fixed length scales absent in the pure Yang-Mills theory. In the Standard Model, the sphaleron is the static, unstable saddle-point solution of the classical field
equations~\cite{Klinkhamer1984,Manton2019}, situated at the energy barrier
separating topologically distinct vacua playing a central role in baryon number violation in the early universe~\cite{KlinkhamerManton1984,MatchevVerner2025}.

The bosonic sector of the electroweak Lagrangian in the limit $\theta_W \to 0$ is
\begin{equation}
\mathcal{L} = -\frac{1}{4}W^a_{\mu\nu}W^{a\mu\nu} + |D_\mu\Phi|^2 - \frac{\lambda}{4}(|\Phi|^2 - v^2)^2,
\label{eq:sphaleron_lagrangian}
\end{equation}
where $\Phi$ is the Higgs doublet and $W^a_{\mu\nu}$ is the SU(2) field strength.

For the spherically symmetric ansatz with gauge profiles $f_A$, $f_B$ and Higgs profiles $H$, $K$~\cite{MatchevVerner2025}, the energy functional becomes
\begin{align}
E_{\rm sph} &= \frac{4\pi}{g^2} \int_0^\infty \mathcal{G}_\rho \, d\rho,
\label{eq:sphaleron_energy}
\end{align}
where $\rho$ is scaled by $m_W$ and
\begin{align}
\mathcal{G}_\rho &= (\dot{f}_A)^2 + (\dot{f}_B)^2 + \frac{(f_A^2 + f_B^2 - 1)^2}{2\rho^2} + 2m_W^2 \rho^2 (\dot{H}^2 + \dot{K}^2) \nonumber \\
&\quad + m_W^2 U_{\rm int} + \frac{m_W^2 m_H^2}{2} \rho^2 (H^2 + K^2 - 1)^2,
\label{eq:sphaleron_G}
\end{align}
with the interaction potential $U_{\rm int} = (f_A^2 + f_B^2)(H^2 + K^2) + 2(f_A H + f_B K)^2$.

The auxiliary quantity is
\begin{align}
\mathcal{C}_\rho &= (\dot{f}_A)^2 + (\dot{f}_B)^2 + 2m_W^2\rho^2(\dot{H}^2 + \dot{K}^2) - \frac{(f_A^2 + f_B^2 - 1)^2}{2\rho^2} - m_W^2 U_{\rm int} \nonumber \\
&\quad - \frac{m_W^2 m_H^2}{2}\rho^2(H^2 + K^2 - 1)^2.
\label{eq:sphaleron_C}
\end{align}
Unlike the BPST instanton where conformal invariance rendered the $\alpha = 1$ identity trivial, here the Higgs mass $m_H$ explicitly breaks scale invariance: the $\rho^2$ factors in the Higgs kinetic and vacuum terms introduce a definite length scale.

Defining the $\alpha$-weighted integrals:
\begin{align}
I_{\rm GK}^\alpha &= \int_0^\infty [(\dot{f}_A)^2 + (\dot{f}_B)^2] \rho^{\alpha-1}\, d\rho, \\
I_{\rm HK}^\alpha &= \int_0^\infty (\dot{H}^2 + \dot{K}^2) \rho^{\alpha+1}\, d\rho, \\
I_{\rm GC}^\alpha &= \int_0^\infty (f_A^2 + f_B^2 - 1)^2 \rho^{\alpha-3}\, d\rho, \\
I_{\rm int}^\alpha &= \int_0^\infty U_{\rm int} \rho^{\alpha-1}\, d\rho, \\
I_{\rm vac}^\alpha &= \int_0^\infty (H^2 + K^2 - 1)^2 \rho^{\alpha+1}\, d\rho,
\end{align}
the virial identity becomes:
\begin{equation}
\alpha I_{\rm GK}^\alpha + 2m_W^2(\alpha - 2) I_{\rm HK}^\alpha = \frac{\alpha - 2}{2} I_{\rm GC}^\alpha + m_W^2 \alpha I_{\rm int}^\alpha + \frac{m_W^2 m_H^2 (\alpha + 2)}{2} I_{\rm vac}^\alpha.
\label{eq:sphaleron_virial}
\end{equation}

The structure of~\eqref{eq:sphaleron_virial} shows how different terms dominate at different $\alpha$ values, each serving as a distinct probe of the sphaleron structure.

For $\alpha = 0$, the gauge kinetic and interaction terms vanish, leaving
\begin{equation}
I_{\rm GC}^0 = 4m_W^2 I_{\rm HK}^0 + m_W^2 m_H^2 I_{\rm vac}^0.
\end{equation}
This relation governs the innermost structure: the centrifugal barrier from topological winding (the $(f_A^2 + f_B^2 - 1)^2/\rho^2$ term, which would diverge without the regularity condition $f_A^2(0) + f_B^2(0) = 1$) is balanced by Higgs gradient pressure and vacuum compression. The $\alpha = 0$ identity is most sensitive to whether the numerical solution correctly implements the regularity condition at the core.

For $\alpha = 2$, the Higgs kinetic and centrifugal terms decouple:
\begin{equation}
I_{\rm GK}^2 = m_W^2 I_{\rm int}^2 + m_W^2 m_H^2 I_{\rm vac}^2.
\end{equation}
Here gauge field gradients are balanced by gauge-Higgs interaction and Higgs potential. The weighting probes the transition region where the sphaleron interpolates to the vacuum.

For $\alpha = 1$
\begin{equation}
I_{\rm GK}^1 - 2m_W^2 I_{\rm HK}^1 = -\frac{1}{2} I_{\rm GC}^1 + m_W^2 I_{\rm int}^1 + \frac{3m_W^2 m_H^2}{2} I_{\rm vac}^1.
\end{equation}
Unlike the BPST case, this identity is \emph{nontrivial}. The Higgs mass explicitly breaks scale invariance, and the coefficient $3m_W^2 m_H^2/2$ of the vacuum term shows that Higgs self-coupling provides additional compressive force proportional to $m_H^2$. For the physical ratio $m_H/m_W \approx 1.55$, this enhancement factor is $(m_H/m_W)^2 \approx 2.4$, so vacuum compression significantly exceeds naive expectations from the gauge sector alone.

The sphaleron thus shows how the $\alpha$-family disentangles contributions from different mechanisms operating at different length scales. A numerical solution that passes the $\alpha = 1$ test may still fail at $\alpha = 0$ if the core regularity condition is poorly resolved, or at $\alpha = 2$ if the vacuum approach is inaccurate.

The sphaleron has four profile functions with specific regularity conditions at the origin and exponential decay at infinity.

\noindent\textit{At the origin} ($\rho \to 0$): The regularity condition $f_A^2(0) + f_B^2(0) = 1$ (typically $f_A(0) = 1$, $f_B(0) = 0$) implies
\begin{equation}
f_A^2 + f_B^2 - 1 \sim \rho^2, \quad H, K \sim \rho.
\end{equation}
The gauge kinetic terms $(\dot{f}_A)^2 + (\dot{f}_B)^2$ remain $O(1)$ at the origin. The most restrictive integral is $I_{\rm GC}^\alpha$:
\begin{equation}
\int (f_A^2 + f_B^2 - 1)^2 \rho^{\alpha-3}\, d\rho \sim \int \rho^4 \cdot \rho^{\alpha-3}\, d\rho = \int \rho^{\alpha+1}\, d\rho,
\end{equation}
which converges for $\alpha > -2$.

\noindent\textit{At infinity} ($\rho \to \infty$): All fields approach their vacuum values exponentially with characteristic scale $m_W^{-1}$, ensuring convergence for all finite $\alpha$.

Then virial relations hold for $\alpha > -2$, with no upper bound.

\section{Skyrmions}
\label{sec:chiral}

The Skyrme model provides an effective field theory of pions whose topological solitons, Skyrmions, model baryons in the large-$N_c$ limit of QCD~\cite{Skyrme1962,AdkinsNappiWitten1983}. The Lagrangian is
\begin{equation}
\mathcal{L} = \frac{f_\pi^2}{16}\text{Tr}(\partial_\mu U^\dagger \partial^\mu U) + \frac{1}{32e^2}\text{Tr}([U^\dagger\partial_\mu U, U^\dagger\partial_\nu U]^2),
\label{eq:skyrme_lagrangian}
\end{equation}
where $U \in \text{SU}(2)$ is the chiral field, $f_\pi$ is the pion decay constant, and $e$ is the Skyrme parameter.

The fundamental $B = 1$ Skyrmion has a spherically symmetric hedgehog ansatz
\begin{equation}
U = \exp(i\bm{\tau}\cdot\hat{\bm{r}}f(\rho)) = \cos f + i\bm{\tau}\cdot\hat{\bm{r}}\sin f,
\label{eq:skyrmion_ansatz}
\end{equation}
where $\bm{\tau}$ are Pauli matrices, $\hat{\bm{r}} = \bm{r}/\rho$, and the profile satisfies $f(0) = \pi$ (baryon number at the core) and $f(\infty) = 0$ (vacuum at infinity).

Unlike the gauge-Higgs systems of the previous section, the Skyrme model requires a term quartic in derivatives for soliton stability. Derrick's theorem forbids static solitons when all terms scale identically under dilation; the Skyrme term, with its different scaling dimension, provides the resistance to collapse that allows equilibrium. Manton~\cite{Manton2024} recently demonstrated that the hedgehog profile varies only weakly across different chiral effective field theories. The virial identities developed here provide a systematic understanding of this robustness: at special $\alpha$ values, different terms decouple, revealing which contributions dominate in different radial regions.

Following Manton's notation, the static energy takes the form
\begin{equation}
E = 4\pi \int_0^\infty \mathcal{G}_\rho \, d\rho, \qquad \mathcal{G}_\rho = \rho^2 \mathcal{E}(f, \dot{f}),
\label{eq:skyrmion_energy_general}
\end{equation}
where the energy density is expanded in symmetrized monomials:
\begin{equation}
\mathcal{E} = \sum_{m,n,p} c_{m,n,p} K_{m,n,p}, \qquad K_{m,n,p} = \sum_{(i,j,k) \in \text{cyc}(m,n,p)} \dot{f}^{\,i} \left(\frac{\sin f}{\rho}\right)^{j+k}.
\label{eq:manton_expansion}
\end{equation}
The index $i$ counts powers of the radial strain $\dot{f}$ (how fast the profile varies radially), while $j+k$ counts powers of the angular strain $\sin f/\rho$ (how much the hedgehog structure costs in angular gradient energy). The competition between radial and angular strain, analogous to the kinetic-potential competition in simpler theories, determines the Skyrmion profile.

\subsection{Virial identity}

Applying the general formalism, each cyclic permutation $(i,j,k)$ contributes to the virial identity with coefficient
\begin{equation}
\alpha(i-1) + (j+k) - 2.
\label{eq:skyrmion_coefficient}
\end{equation}
The factor $(i-1)$ arises from Euler's theorem applied to terms of degree $i$ in $\dot{f}$; the factor $(j+k) - 2$ arises from the explicit $\rho$-dependence of the angular strain terms. The general virial identity for arbitrary chiral effective Lagrangians is
\begin{equation}
\sum_{m,n,p} c_{m,n,p} \sum_{(i,j,k) \in \text{cyc}(m,n,p)} \left[\alpha(i-1) + (j+k) - 2\right] I_{i,j,k}^\alpha = 0,
\label{eq:skyrmion_virial_general}
\end{equation}
where $I_{i,j,k}^\alpha = \int_0^\infty \dot{f}^{\,i} (\sin f)^{j+k} \rho^{\alpha+1-j-k}\, d\rho$.

\subsection{Standard Skyrme model}

The standard model has $c_{2,0,0} = c_{2,2,0} = 1$, giving the reduced functional
\begin{equation}
\mathcal{G}_\rho = \rho^2 \dot{f}^2 + 2\sin^2 f + 2\dot{f}^2 \sin^2 f + \frac{\sin^4 f}{\rho^2},
\label{eq:skyrmion_G}
\end{equation}
so that $E = 4\pi \int_0^\infty \mathcal{G}_\rho \, d\rho$. The four terms represent: Dirichlet (radial) gradient energy $\rho^2\dot{f}^2$; sigma-model (angular) gradient energy $2\sin^2 f$; Skyrme term $2\dot{f}^2\sin^2 f$ mixing radial and angular strains; and centrifugal barrier $\sin^4 f/\rho^2$ that prevents collapse.

The auxiliary quantity is
\begin{equation}
\mathcal{C}_\rho = \rho^2\dot{f}^2 + 2\dot{f}^2\sin^2 f - 2\sin^2 f - \frac{\sin^4 f}{\rho^2}.
\label{eq:skyrmion_C}
\end{equation}

The virial identity becomes 
\begin{equation}
(\alpha - 2) I_D^\alpha + 2\alpha I_S^\alpha = 2\alpha I_\sigma^\alpha + (\alpha - 2) I_C^\alpha,
\label{eq:skyrmion_virial}
\end{equation}
where $I_D^\alpha$, $I_S^\alpha$, $I_\sigma^\alpha$, $I_C^\alpha$ are the $\alpha$-weighted integrals of the Dirichlet, Skyrme, sigma-model, and centrifugal terms respectively.

For the standard Skyrme model, the identity~\eqref{eq:skyrmion_virial} was first obtained
by Gudnason, Gao, and Yang~\cite{GudnasonGaoYang2017} (their Eq.~(4.30) with
$m=0$ and $2\kappa=\alpha+2$ by using . We rederive it here within our formalism
and discuss its special $\alpha$ values.
The virial identity~\eqref{eq:skyrmion_virial} simplifies at particular $\alpha$ values, isolating specific balances:

For $\alpha = 1$, the identity reduces to $E_2 = E_4$, the Derrick relation stating that total derivative-squared energy equals total derivative-fourth energy. This is the classical result explaining Skyrmion stability.

For $\alpha = 0$, we obtain $I_D^0 = I_C^0$: the Dirichlet gradient weighted by $\rho$ balances the centrifugal barrier weighted by $\rho^{-3}$. Both integrands scale as $\rho$ for a regular profile $f = \pi - \beta\rho + O(\rho^3)$, making this identity most sensitive to core structure.

For $\alpha = 2$, the Dirichlet and centrifugal terms drop out, leaving $I_S^2 = I_\sigma^2$: the $\rho$-weighted average of $\dot{f}^2$ over the angular strain density equals unity. This constrains the intermediate region where $\sin f$ is largest, the ``waist'' of the Skyrmion where the hedgehog structure transitions from core to tail.

Manton notes that $K_{2,0,0}$, $K_{2,2,0}$, and $K_{2,2,2}$ are the only symmetric polynomials quadratic in time-derivatives~\cite{Manton2024}. The sextic contribution $K_{2,2,2} = 3\dot{f}^2 \sin^4 f / \rho^4$ has coefficient $\alpha + 2$ in the virial identity, yielding
\begin{equation}
(\alpha - 2) I_D^\alpha + 2\alpha I_S^\alpha + 3\nu_6(\alpha + 2) I_6^\alpha = 2\alpha I_\sigma^\alpha + (\alpha - 2) I_C^\alpha.
\end{equation}

At $\alpha = -2$, the BPS-Skyrme term decouples: $I_D^{-2} + I_S^{-2} = I_\sigma^{-2} + I_C^{-2}$. This probes the core independently of the sextic coupling $\nu_6$, useful for testing numerical implementations where the BPS-Skyrme term might introduce additional discretization errors.

\subsection{Asymptotic universality}

For large $\rho$, the profile approaches $f \sim C/\rho^2$. The coefficient $(2-j-k)$ in the $\rho$-derivative of the integrand controls the asymptotic falloff: the Dirichlet term ($j+k=0$) dominates as $\rho^{-5}$, while terms with larger $j+k$ decay faster. Consequently,
\begin{equation}
\frac{d\mathcal{C}_\rho}{d\rho} \approx -2\rho \dot{f}^2 \quad (\rho \to \infty).
\end{equation}
This confirms Manton's observation~\cite{Manton2024} that the tail is determined entirely by the Dirichlet term, with the angular-strain terms becoming negligible at large $\rho$. The virial formalism provides a systematic derivation of this asymptotic universality.

The Skyrmion has power-law decay at infinity (massless pions), yielding a finite $\alpha$ range.

\noindent\textit{At the origin} ($\rho \to 0$): The hedgehog boundary condition $f(0) = \pi$ and regularity require
\begin{equation}
f = \pi - \beta\rho + O(\rho^3), \quad \dot{f} \sim -\beta, \quad \sin f = \sin(\pi - \beta\rho) \sim \beta\rho.
\end{equation}
Each term in $\mathcal{C}_\rho$ scales as $\rho^2$:
\begin{equation}
\rho^2\dot{f}^2 \sim \rho^2, \quad \dot{f}^2\sin^2 f \sim \rho^2, \quad \sin^2 f \sim \rho^2, \quad \frac{\sin^4 f}{\rho^2} \sim \rho^2.
\end{equation}
Thus $\mathcal{C}_\rho \sim \rho^2$ and $\rho^\alpha \mathcal{C}_\rho \sim \rho^{\alpha+2} \to 0$ for $\alpha > -2$.

\noindent\textit{At infinity} ($\rho \to \infty$): The massless pion field decays as a power law,
\begin{equation}
f \sim \frac{C}{\rho^2}, \quad \dot{f} \sim -\frac{2C}{\rho^3}, \quad \sin f \approx f \sim \rho^{-2}.
\end{equation}
The Dirichlet term dominates: $\rho^2\dot{f}^2 \sim \rho^{-4}$. The angular terms decay faster: $\sin^2 f \sim \rho^{-4}$, $\dot{f}^2\sin^2 f \sim \rho^{-10}$, $\sin^4 f/\rho^2 \sim \rho^{-10}$. Thus $\mathcal{C}_\rho \sim \rho^{-4}$ and
\begin{equation}
\rho^\alpha \mathcal{C}_\rho \sim \rho^{\alpha-4} \to 0 \quad \text{for } \alpha < 4.
\end{equation}

As a consequence the valid range is $-2 < \alpha < 4$. The finite upper bound reflects the power-law decay of massless pions.

\section{Conclusion}
\label{sec:conclusion}

We have derived a continuous family of virial identities for O($n$) symmetric field configurations, parameterized by an exponent $\alpha$ that controls the radial weighting. The family extends the classical Derrick constraint ($\alpha = 1$) to a set of identities that probe different radial regions: negative $\alpha$ emphasizes the core where field gradients are steepest, large positive $\alpha$ emphasizes the asymptotic tail, and $\alpha = 1$ samples the profile uniformly.

The numerical examples illustrate how the $\alpha$-dependence of errors distinguishes between core and tail inaccuracies. For Nielsen-Olesen vortices, the errors grow at negative $\alpha$: the $\alpha = 1$ identity held to $0.0005\%$ while $\alpha = -0.5$ showed a $5.7\%$ discrepancy, exposing core-region errors that the global Derrick test averages away. For the Bounce solutions, the opposite pattern emerged: errors grew monotonically with $\alpha$, from $0.001\%$ at $\alpha = -2$ to $0.03\%$ at $\alpha = 2$, indicating that numerical inaccuracies were concentrated in the asymptotic tail where the bounce approaches the false vacuum. The $\alpha$-family thus gives a clear signature: core errors manifest at negative $\alpha$, while tail errors manifest at positive $\alpha$.

For BPS configurations (the monopole, BPST instanton, and critical vortex), the first-order Bogomolny equations imply pointwise equality between kinetic and potential densities, guaranteeing automatic satisfaction of the virial identity for all valid $\alpha$. A numerical BPS solution that satisfies some $\alpha$ values but not others has failed to solve the Bogomolny equations accurately in the corresponding radial region.

Analytical verification was provided by the Fubini-Lipatov instanton, where conformal invariance renders the $\alpha = 1$ test trivial (satisfied for any instanton size $R$), while the $\alpha \neq 1$ relations constrain the action density distribution. The electroweak sphaleron identity displays the interplay between gauge and Higgs sectors, with the Higgs mass explicitly breaking the scale invariance that trivializes the pure Yang-Mills case. The $\alpha$-dependent structure shows how different terms dominate at different radial scales: the centrifugal barrier and regularity condition at $\alpha = 0$, the gauge-Higgs transition at $\alpha = 2$, and the enhanced vacuum compression from the Higgs self-coupling at $\alpha = 1$. The Skyrmion identity handles arbitrary chiral effective Lagrangian terms, with special $\alpha$ values isolating specific contributions: $\alpha = 0$ probing the core, $\alpha = 2$ probing the intermediate region, and $\alpha = -2$ decoupling the BPS-Skyrme term.

Natural generalizations of the present formalism include the extension to theories with fermionic fields, where soliton zero modes may satisfy analogous radial constraints, and to finite-temperature configurations, where compactification of Euclidean time breaks O(4) symmetry and introduces calorons and thermal sphalerons as the relevant objects.

\acknowledgments

The author thanks Manuel Torres Labansat for discussions on vacuum decay and bounce solutions.

\bibliographystyle{JHEP}
\bibliography{apssamp}

\end{document}